\documentclass[%
 reprint,onecolumn,
superscriptaddress,
nofootinbib,
 pre,amsmath,amssymb,
]{revtex4-2}

\usepackage{url}
\usepackage{graphicx}% Include figure files
\usepackage{bm}% bold math
\usepackage[dvipsnames]{xcolor}
\usepackage{multirow}

\DeclareMathOperator{\EX}{\mathbb{E}}% expected value

\begin{document}
\title{Ergodicity breaking in wealth dynamics: The case of reallocating geometric Brownian motion}
\author{Viktor Stojkoski}
\thanks{Corresponding author}
\email{vstojkoski@eccf.ukim.edu.mk}
\affiliation{Faculty of Economics, Ss.~Cyril and Methodius University, 1000 Skopje, Macedonia}
\affiliation{Research Center for Computer Science and Information Technologies, Macedonian Academy of Sciences and Arts, Bul. Krste Misirkov 2, 1000 Skopje, Macedonia}

\author{Marko Karbevski}
\email{markokarbevski@gmail.com}
\affiliation{Polar Cape Consulting, Sankt Eriksgatan 63b 11234 Stockholm, Sweden}
\affiliation{Sorbonne Université, 4 Place Jussieu, 75005 Paris, France}
\affiliation{Institute of Mathematics, Faculty of Natural Sciences and Mathematics, Ss.~Cyril and Methodius University, Arhimedova 3,
1000 Skopje, Macedonia}
\affiliation{Research Center for Computer Science and Information Technologies, Macedonian Academy of Sciences and Arts, Bul. Krste Misirkov 2, 1000 Skopje, Macedonia}

\date{\today}

\begin{abstract}
A growing body of empirical evidence suggests that the dynamics of wealth within a population tends to be non-ergodic, even after rescaling the individual wealth with the population average. Despite these discoveries, the way in which non-ergodicity manifests itself in models of economic interactions remains an open issue. Here, we shed valuable insight on these properties by studying the non-ergodicity of the population average wealth in a simple model for wealth dynamics in a growing and reallocating economy called Reallocating geometric Brownian motion (RGBM). When the effective wealth reallocation in the economy is from the poor to the rich, the model allows for the existence of negative wealth within the population. We show that then, in RGBM ergodicity breaks as the difference between the time-average and the ensemble growth rate of the average wealth in the population. In particular, the ensemble average wealth grows exponentially whereas the time-average growth rate is non-existent. Moreover, we find that the system is characterized with a critical self-averaging time period. Before this time period, the ensemble average is a fair approximation for the population average wealth. Afterwards, the non-ergodicity forces the population average to oscilate between positive and negative values since then the magnitude of this observable is determined by the most extreme wealth values in the population. This implies that the dynamics of the population average is an unstable phenomenon in a non-ergodic economy.  We use this result to argue that one should be cautious when interpreting economic well-being measures that are based on the population average wealth in non-ergodic economies.
\end{abstract}

\maketitle

\section{Introduction}

Mathematical models of economies are often concerned with the dynamics of the wealth within a population. By definition, wealth is a growing quantity, and hence it is non-ergodic~\cite{peters2018ergodicity}. The non-ergodicity implies that ensemble values are not an adequate representation for the behavior of the economy over time. A standard approach for dealing with this problem is to transform the wealth $x_i$ of each person $i$ by dividing it with the  population average, $\langle x \rangle_N  = \sum_i x_i / N$, and then assume that the resulting rescaled wealth is ergodic~\cite{gabaix2016dynamics}. This allows the implementation of powerful mathematical tools for studying the evolution of various economic phenomena which are a result of the wealth dynamics. For instance, economic inequality and social mobility measures intuitively include rescaling in their definitions for the purpose of being used for comparisons across economies and between periods~\cite{burkhauser2009intragenerational,chakrabarti2013econophysics,jantti2015income,saez2016wealth}. 

However, recent empirical investigations suggest that even the rescaled wealth might be a non-ergodic observable. In particular, by utilizing a model of economic interactions called Reallocating geometric Brownian motion (RGBM), and data for the United States, Berman et al.~\cite{berman2017empirical} showed that the dynamics of rescaled wealth over time varies between phases of ergodic and non-ergodic behavior. RGBM is a simple model of a closed one-generation economy that allows to capture the possible non-ergodic dynamics within the system~\cite{bouchaud2000wealth,liu2017correlation,marsili1998dynamical,furioli2017fokker}. The model distinguishes three regimes depending on the orientation of the wealth reallocation in the economy. In the first, i.e. the positive regime where the reallocation is from the rich to the poor, rescaled wealth is an ergodic quantity and it is positive for each individual. The no-reallocation regime reduces to independent geometric Brownian motion (GBM) trajectories~\cite{stojkoski2020generalised}. In this case, it is known that the system is non-ergodic, and eventually one person ends up owning all the wealth in the economy~\cite{peters2013ergodicity}. Finally, in the negative reallocation regime the reallocation of wealth is from the poor to the rich. Besides the non-ergodic dynamics, this regime further allows the existence of negative wealth among the individuals. A distinct characteristic of the negative reallocation regime is the presence of a self-averaging time period during which economic inequality increases whereas social mobility decreases, consistent with current empirical observations~\cite{hurst1998wealth,klevmarken2003wealth,kennickell2011tossed}. Strikingly, Berman et al.~\cite{berman2017empirical} found out that the current wealth dynamics in the United States are best described with this regime.

Since the negative reallocation regime has only recently been discovered as a plausible explanation for realistic wealth dynamics, the properties of RGBM in this regime remain an open issue. In the absence of a theoretical background, research has focused on the properties of RGBM in infinite populations for which ensemble estimates are a good approximation~\cite{peters2018sum}. While this is a fairly good assumption in a large population, in reality every population is finite and eventually it will not follow the ensemble behavior. This raises the important question of what happens to the population average wealth in a non-ergodic economy as time goes towards infinity.

In this paper, we provide mathematical reasoning for the non-ergodicity in the negative reallocation regime of RGBM. We show analytically and display numerically that the non-ergodic behavior is manifested in the difference between the ensemble and time-average growth rate of the population average wealth. The ensemble average grows with an exponential rate, whereas the time-average growth rate is non-existent.  Moreover, we find that there is a self-averaging time period. During this period, the ensemble average is a good approximation of the population behavior. After that, the population average wealth oscillates non-regularly around two symmetrical boundary values. The magnitude of the boundary values implies that after self-averaging we are going to observe phases in which the observed growth is larger than the ensemble prediction, but also phases in which the population average wealth is negative. As a consequence, the standard rescaled wealth may appear as an ill-defined quantity and, obviously, dividing the individual wealth with the population average cannot be used as an ergodic transformation. This opens up a discussion on how the dynamics of wealth should be compared in non-ergodic economies in order to derive efficient economic policies. We emphasize that RGBM and its extensions have a long history in the statistical physics community. The model is also known as the Bouchaud-Mezard model for economic exchanges~\cite{bouchaud2000wealth,marsili1998dynamical,liu2017correlation,gueudre2014explore}, and its applications vary from explaining evolution of cooperation~\cite{peters2015evolutionary,stojkoski2019cooperation,stojkoski2021evolution}, up to describing ontogenetic mass properties~\cite{west2012allometry,holden2013change}. Thus, our results may also induce important implications to the long time behavior of multiple social and natural phenomena.

The rest of the paper is organized as follows. In Section~\ref{sec:preliminaries} we describe RGBM and its known properties. In Section~\ref{sec:results} we present our analysis for the non-ergodicity of the model. The last section discusses the implications created by our results.

\section{RGBM as a model of wealth dynamics}
\label{sec:preliminaries}

Under RGBM, the dynamics of the wealth $x_i(t)$ of each individual $i$ at time $t$ is specified as
\begin{align}
    \mathrm{d} x_i &= x_i \left( \mu \mathrm{d}t + \sigma \mathrm{d}W_i \right) - \tau \left( x_i - \langle x \rangle_N \right) \mathrm{d}t,
\label{eq:rgbm}
\end{align}
with $\mu$ being the drift term, $\sigma > 0$ the noise amplitude, and $\mathrm{d}W_i$ is an independent Wiener increment, $W_i(t) =\int_0^t \mathrm{d}W_i$. We assume that the initial values $x_i(0)$ are identically distributed with mean $x_0$, variance $v_0 - x_0^2$ and a covariance $r_0 - x_0^2$.

In the equation $\tau$ is a parameter that quantifies reallocation of wealth. The parameter implies that every year, everyone in the economy contributes a proportion $\tau$ of their wealth to a central pool, and then the pool is shared evenly across the population. The parameter aggregates a multitude of effects: collective investment in infrastructure, education, social programs, taxation, rents paid, private profits made etc. 

Notice that the model does not account for a large amount of important characteristics that may constitute an economy, such as openness of the economy (trade with other economies), intergenerationality (interactions between different generations) or direct effects of economic policies. Instead, it focuses solely on the wealth dynamics that are due to individual growth and a consequence of interactions between the individuals. While this may be seen as a drawback of the model, we believe that in fact it is the major advantage of RGBM. It allows us to isolate the effect of non-ergodic wealth dynamics and investigate the resulting implications.  

The mathematical properties of RGBM when $\tau \geq 0$ are known~\cite{bouchaud2000wealth,liu2017correlation}. In particular, for $\tau > 0$ the growth rate of the population average wealth is an ergodic observable and the model exhibits mean-reversion. That is, each $x_i$ eventually reverts to the population average $\langle x \rangle_N$. The large population approximation for the population average wealth $\langle x(t) \rangle_N = \exp \left[\mu t\right]$ is valid, and rescaled wealth $y_i = x_i / \langle x \rangle_N$ has a stationary probability distribution, %The stationary distribution reads
\begin{align}
    p(y) &= \frac{(\theta - 1)^{\theta}}{\Gamma(\theta)} \exp{\big(-\frac{\theta - 1}{y}\big)} y^{-(1+\theta)},
    \label{eq:stationary-distributin}
\end{align}
where $\theta = 1 + \frac{2 \tau}{\sigma^2}$ and $\Gamma(\cdot)$ is the Gamma function.
This distribution exhibits a power law tail and in probability theory is known as the Inverse gamma distribution.

The ergodicity of the population average wealth allows us to use the stationary distribution in order analytically quantify standard indices for economic well-being and subsequently use them to derive economic policies. For instance, the Gini coefficient of the stationary distribution can be used as a measure of economic inequality~\cite{stojkoski2021income}. The expression for the Gini index is is
\begin{align}
    G &= \int_0^\infty F(y) \left(1 - F(y) \right) dy,
    \label{eq:rgbm-gini}
\end{align}
where $F(y) = \int_0^{y}  p(z) dz = \Gamma(\theta,\frac{\theta-1}{y})$ is the cumulative distribution function of the stationary distribution.
Formally, economic inequality is defined as the extent of concentration in the distribution of wealth among the population. In this aspect, a higher Gini coefficient implies that the total wealth in the economy is concentrated in few individuals, i.e., the society is more unequal. It can be shown that the solution to Eq.~\eqref{eq:rgbm-gini} is a decreasing function with respect to $\tau$ and an increasing with respect to $\sigma$. Thus, RGBM predicts that the inequality in the economy can be reduced by increasing the rate of reallocation, or reducing the impact of randomness (reducing $\sigma$).

Without reallocation ($\tau = 0$), the model is just GBM. Under GBM wealth is non-ergodic and it follows a lognormal distribution which broadens indefinitely over time. There is no stationary non-zero distribution to which rescaled wealth converges. In GBM inequality is always increasing and mobility is always decreasing.

As pointed out, recent empirical evidence suggests that we are currently living in a negative $\tau$ regime~\cite{berman2017empirical}. Not much is known about this regime except that there is a self-averaging time period during which individual trajectories repel from the population mean. This introduces negative individual wealth, a phenomenon observed in almost every modern economy, and makes the system non-ergodic.
The properties of the model after the self-averaging period are unknown. In what follows, we examine these properties from both an analytical and numerical perspective.

\section{Non-ergodicity in RGBM}
\label{sec:results}

\begin{figure*}[t!]
\includegraphics[width=17cm]{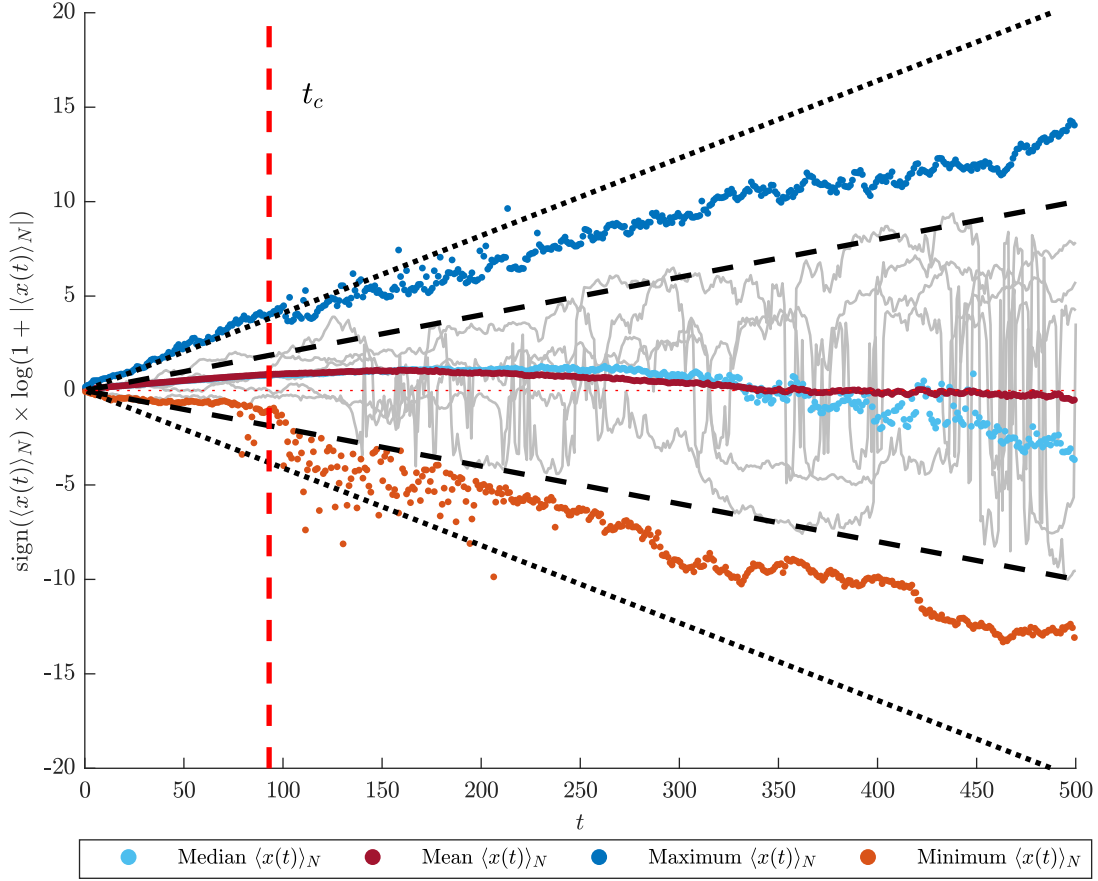}
\caption{RGBM negative $\tau$ regime behavior. \underline{\textbf{Scatter lines:}} Numerical estimations for the median, mean maximum and minimum of the wealth $\langle x(t) \rangle_N $ averaged across $10^3$ simulations of RGBM. \underline{\textbf{Dashed line:}} Exponential growth with rate $\mu$. \underline{\textbf{Dotted line:}} Exponential growth with rate $\mu-\tau +\frac{\sigma^2}{2}$. \underline{\textbf{Vertical line:}} Critical self-averaging time $t_c$. \underline{\textbf{Parameters:}} We set $\mu = 0.021$, $\sigma^2 = 0.02$, $\tau = -0.01$ and $N = 10$.  The initial condition $x_i(0) = 1$ for all $i$.\label{fig:rgbm-behavior}}
\end{figure*}

The non-ergodic behavior of the population average wealth in RGBM is summarized in 
Fig.~\ref{fig:rgbm-behavior}. The figure illustrates the dynamics of the population average $\langle x(t) \rangle_N $ when $\tau$ is negative. To calculate the typical RGBM dynamics we average across $10^3$ simulations of RGBM, as is done in practice~\cite{peters2013ergodicity,peters2018sum,stojkoski2021autocorrelation,stojkoski2021geometric}. We observe two distinct behaviors in the dynamics which are divided by a critical self-averaging time $t_c$ (the red vertical line). The time period when $t < t_c$ is the self-averaging period. During this period the ensemble average is a good approximation for the population behavior, $\langle x(t) \rangle_N \sim \exp\left[ \mu t\right]$. Afterwards, the non-ergodicity forces the population average to be dominated by extreme values, which can be both positive and negative, due to the existence of individuals with negative wealth. The gray lines in the background of the figure show samples of trajectories of the single simulation runs. It is evident that after the self-averaging period we observe complex oscillatory behavior in the population average. This behavior is characterized with envelopes (dashed black lines) which determine the magnitude of fluctuations of $\langle x(t) \rangle_N $. Let us proceed with formal proofs for the long-time properties of the population average wealth.

\subsection{Time average and ensemble growth rate of the population average wealth} 

The non-ergodicity depicted in Fig.~\ref{fig:rgbm-behavior} is a result of the difference between the ensemble and time-average \textit{growth rate} of the population average wealth. This growth rate is defined as
\begin{align}
    g(t,N) &= \frac{1}{t}\log \left(\langle x(t) \rangle_N \right).
    \label{eq:growth-rate}
\end{align}
The ensemble growth rate is found by fixing the period $t$ and taking the limit as the population size grows infinitely. The solution is
\begin{align}
  \lim_{N\to \infty}  g(t,N) &= \mu.
    \label{eq:ensemble-growth-rate}
\end{align}
This will be shown subsequently using It\^{o}'s Lemma. 
On the other hand, the time-average growth rate is found by fixing the population size $N$ and letting time remove the stochasticity. This limit is non-existent. 
Let us study the non-comutativity in the two limits of Eq.~(\ref{eq:growth-rate}). 

\paragraph{Ito Lemma for RGBM:} The ensemble growth rate can be found by studying It\^{o}'s Lemma in the case of averages of autonomous functions of RGBM.
In general, the lemma states that the differential of an arbitrary one-dimensional twice-differentiable function $f(\mathbf{x},t)$ governed by an It\^{o} drift-diffusion process (such as equation~(\ref{eq:rgbm})), is given by
\begin{equation}\begin{aligned}
\mathrm{d}f(\mathbf{x},t) = &\frac{\partial f}{\partial t} \mathrm{d}t + \sum_i \frac{\partial f}{\partial x_i}\mathrm{d}x_i + \frac{1}{2}\sum_i \sum_j \frac{\partial^2 f}{\partial x_i \partial x_j} \mathrm{d}x_i \mathrm{d}x_j. 
\label{eq:ito-lemma}
\end{aligned}
\end{equation}
From~(\ref{eq:ito-lemma}) the general It\^{o} formula for averages of autonomous functions is
\begin{equation}
\begin{aligned}
\mathrm{d}\EX \left[f(\mathbf{x}) \right] =  &\sum_i \EX \left[\frac{\partial f}{\partial x_i}  \mathrm{d}x_i \right] + \frac{1}{2} \sum_i \sum_j \EX \left[ \frac{\partial^2 f}{\partial x_i \partial x_j} \mathrm{d}x_i \mathrm{d}x_j \right]
\label{eq:ito-lemma-expectation}
\end{aligned}
\end{equation}
where we introduce the notation $$\EX \left[f(\mathbf{x}) \right] = \lim_{N\to\infty} = \langle f(\mathbf{x}) \rangle_N$$ as a means to differentiate between ensemble average and a finite sample average of size $N$.

For RGBM we can utilize the independent Wiener increment property $\langle \mathrm{d}W_i^2 \rangle = \mathrm{d}t$, and omit terms of order $\mathrm{d}t^2$ as they are negligible. Then,
\begin{align}
   \EX \left[ \frac{\partial f}{\partial x_i} \mathrm{d}x_i \right] &= (\mu -\tau)  \EX \left[ \frac{\partial f}{\partial x_i} x_i \right] \mathrm{d}t + \tau \EX \left[ \frac{\partial f}{\partial x_i} \langle x \rangle_N \right] \mathrm{d}t
\label{eq:rgbm-1st-term}
\end{align}
and
\begin{align}
   \EX \left[ \frac{\partial^2 f}{\partial x_i \partial x_j} \mathrm{d}x_i \mathrm{d}x_j \right] &= 
    \begin{cases}
\sigma^2 \EX \left[ \frac{\partial^2 f}{\partial x_i \partial x_j} x_i x_j \right] \mathrm{d}t & \text{if $i = j$,} \\
0 & \text{otherwise.} 
\end{cases}
\label{eq:rgbm-cross-term}
\end{align}
The result in equation~(\ref{eq:rgbm-cross-term}) can be seen by writing out one cross-term of $\mathrm{d}x_i \mathrm{d}x_j$ as
\begin{align*}
     O( \mathrm{d}t^2) + O( \mathrm{d}t \sigma \mathrm{d}W_i) + O( \mathrm{d}t \sigma \mathrm{d}W_j) + x_i x_j \sigma^2 \mathrm{d}W_i \mathrm{d}W_j.
\end{align*}
By inserting the estimates in Eq.~(\ref{eq:ito-lemma-expectation}) we can write the It\^{o} formula for the average of non-autonomous functions in RGBM as
\begin{equation}
\label{eq:ito-lemma-RGBM}
\begin{aligned}
\frac{\mathrm{d}\EX \left[f(\mathbf{x})\right]}{\mathrm{d}t} = &\big( \mu - \tau \big) \sum_i \EX \left[ \frac{\partial f}{\partial x_i} x_i \right]+ \tau \EX \left[ \langle x \rangle_N \sum_i  \frac{\partial f}{\partial x_i} \right] + \frac{\sigma^2}{2}\sum_i\EX \left[ \frac{\partial^2 f}{\partial x_i^2} x_i^2 \right].
\end{aligned}
\end{equation}

\paragraph{Ensemble average growth rate:} To calculate the ensemble growth rate we set $f(\mathbf{x}) = x $. Then, the differential equation which governs the evolution of $\EX\left[x(t) \right]$ is 
\begin{align}
  \frac{\mathrm{d} \EX \left[  x \right]}{\mathrm{d}t} &= \mu \EX \left[ x \right],      \label{eq:rgbm-1st-moment-individual-stochastic-differential}
\end{align}
whose solution is simply
\begin{align}
  \EX \left[ x(t) \right] &=  \EX \left[ x(0) \right] \exp \left[ \mu t \right].
    \label{eq:rgbm-population-expectation}
\end{align}
Inserting Eq.~(\ref{eq:rgbm-population-expectation}) in Eq.~(\ref{eq:growth-rate}) we get that the ensemble growth rate is Eq.~(\ref{eq:ensemble-growth-rate}).

\paragraph{Time average growth rate:} We prove that the time-average growth rate does not exist by using contradiction. That is, we assume that this limit exists and $\lim_{t\to\infty} g(t,N) = \gamma $. Then, for large enough $t$ we can approximate Eq.~(\ref{eq:rgbm}) as
\begin{equation}
\begin{aligned}
    \mathrm{d} x_i &= x_i \left( \left[\mu-\tau\right] \mathrm{d}t + \sigma \mathrm{d}W_i \right) + \tau \exp\left[\gamma t\right] \mathrm{d}t.
\label{eq:rgbm-approximate}
\end{aligned}
\end{equation}
Equation~(\ref{eq:rgbm-approximate}) is a one-dimensional linear stochastic differential equation whose solution reads
\begin{equation}
\begin{aligned}
    &x_i(t) = \exp\left[(\mu-\tau-\frac{\sigma^2}{2})t + \sigma W_i(t) \right] \times \bigg( \tau \int_0^t \exp\left[(\gamma-(\mu-\tau-\frac{\sigma^2}{2}))s - \sigma W_i(s) \right]\mathrm{d}s +x_i(0) \bigg).
    \label{eq:rgbm-approximate-solution}
\end{aligned}
\end{equation}
Let us examine two cases i) $\gamma \geq \mu - \tau - \frac{\sigma^2}{2}$, and ii) $\gamma < \mu - \tau - \frac{\sigma^2}{2}$. In the first case the integral in equation~(\ref{eq:rgbm-approximate-solution}) diverges and, since $\tau < 0$, eventually each $x_i(t)$ becomes negative. Therefore, the solution to $g(t,N)$ is undefined as $t \to \infty$.

In the case when $\gamma < \mu - \tau - \frac{\sigma^2}{2}$ the integral converges to a certain value $c_i$ which may be dependent on the random realization. For a sufficiently large $t$ we can write Eq.~(\ref{eq:rgbm-approximate-solution}) as
\begin{align*}
    x_i(t) &= \exp\left[(\mu-\tau-\frac{\sigma^2}{2})t + \sigma W_i(t) \right] \bigg( x_i(0) + \tau c_i\bigg)
\end{align*}
and the population average as
\begin{align*}
    \langle x(t) \rangle_N &=  \exp\left[(\mu-\tau-\frac{\sigma^2}{2})t  \right] \times \frac{1}{N} \sum_i \exp\left[\sigma W_i(t) \right] (x_i(0) + \tau c_i).
\end{align*}
Next, we define the event $A_i$ as the situation when $c_i > \frac{x_i(0)}{|\tau|}$. The probability $\mathrm{Pr}(A_i)$ is always greater than zero. This follows directly from the properties of Brownian motion. In particular, notice that
\begin{align*}
\mathrm{Pr} \left( \min_{t \in [a,b]} (\kappa t - \sigma W_i(t))  > \epsilon \right) > 0,
\end{align*}
for any constants $\kappa$ and $\epsilon$, and $0<a<b$ ~\cite{Durett}. Exponentiation of the term inside the probability notation yields $\mathrm{Pr}(A_i)>0$.

Let $A$ be the intersection of all $A_i$. Notice that $A$ is included in the event that the limit superior of the average population wealth $\langle x(t)\rangle_N$ is less than zero. Since the Wiener processes $(W_i)_{i \in \{1, \cdots ,N \}}$ are independent, this leads to the independence of $(A_i)_{i \in \{1, \cdots ,N \}}$. Hence,
\begin{align*}
    \mathrm{Pr}\left(\limsup_{t\to\infty} \langle x(t) \rangle_N < 0\right) &\ge \mathrm{Pr}\left(\bigcap_{i=1}^N A_i \right) = \prod_i^N \mathrm{Pr}(A_i) > 0
\end{align*}
for any finite population size $N$.
In words, there is always a positive probability in the time limit that the population average becomes negative. This violates our initial assumption for its limiting growth rate, thus concluding the proof that the time-average growth rate in RGBM is non-existent.

We point out that the same techniques can be used for showing that the population average is not always negative in the time limit. In particular, by assuming that $\langle x(t) \rangle_N = -\exp\left[\gamma t \right]$ and substituting it in Eq.~\eqref{eq:rgbm}, we can arrive at the same contradictory arguments, i.e, there is always a positive probability to observe a positive population average in the time limit. Hence, due to the non-ergodicity, the population average will oscillate within an interval with symmetric boundaries.

\subsection{Critical self-averaging time}

We showed how the population average wealth behaves for finite $N$ as time goes towards infinity. To characterize the behavior of the population average wealth for finite $t$ and fixed $N$ we resort to the concept of self-averaging. In statistical physics self-averaging is known as the situation when a sample average resembles the corresponding ensemble value, i.e., the time until Eq.~(\ref{eq:rgbm-population-expectation}) is valid. A simple strategy for estimating when this occurs is to look at the relative variance of the population average wealth $\langle x(t)\rangle_N$,
\begin{align}
    \mathrm{R}_N(t) &\equiv \frac{\mathrm{var}(\langle x(t) \rangle_N)}{\EX\left[\langle x(t) \rangle_N\right]^2},
    \label{eq:relative-variance}
\end{align}
where $\mathrm{var}(\mathrm{x}) = \EX \left[ \mathrm{x}^2 \right]-\EX \left[ \mathrm{x} \right]^2 $ is the variance of $\mathrm{x}$.

If $\mathrm{R}_N(t)$ converges to $0$ in the time limit, the system is self-averaging and the population average wealth will always resemble the ensemble average. In RGBM this is always true for $\tau \geq \frac{\sigma^2}{2}$. When $\tau < \frac{\sigma^2}{2}$, the system will experience self-averaging until some critical time $t_c$ which is dependent on both the initial condition and the population size $N$. Afterwards it will collapse to its time-average behavior. The critical time can be found by rewriting Eq.~(\ref{eq:relative-variance}) as
\begin{align}
\mathrm{R}_N(t) &= \frac{\EX\left[\langle x(t) \rangle_N^2\right] - \EX\left[\langle x(t) \rangle_N\right]^2}{\EX\left[\langle x(t) \rangle_N\right]^2}, \nonumber \\
&= \frac{1}{N^2} \frac{\EX\left[ \sum_i \sum_{j} x_i(t) x_j(t) \right]}{\EX\left[\langle x(t) \rangle_N\right]^2} -1, \nonumber \\
&=\frac{ \sum_i \sum_{j} \EX\left[  x_i(t) x_j(t) \right]}{ \left( \sum_i \EX\left[ x_i(t) \right] \right)^2} -1 
\end{align}
If $\mathrm{R}(N,t) << 1$, then the population average will likely be close to its ensemble average value. Thus,
the system will be self-averaging until the critical point $t_c$ which occurs at $\mathrm{R}_N(t_c) = 1$.

The difficulty in estimating the relative variance arises because the dynamics of $\EX\left[ x_i(t) x_j(t) \right]$ are coupled for every $i$ and $j$ whenever $\tau \neq 0$, and therefore their evolution is interdependent. This issue can be resolved by interpreting the dynamics of $\EX\left[ x_i(t) x_j(t) \right]$ as a system of differential equations and utilizing the RGBM It\^{o} Lemma. Then, by setting $f(\mathbf{x}) = x_i x_j$ in ~(\ref{eq:ito-lemma-RGBM}), it follows that the dynamics of $\EX\left[x_i(t) x_j(t)\right]$ can be described as
\begin{equation}
\begin{aligned}
    \frac{\mathrm{d}\EX\left[x_i x_j\right]}{\mathrm{d}t} =   \begin{cases}
2 ( \mu - \frac{N-1}{N} \tau + \frac{\sigma^2}{2}) \EX\left[x_i^2\right] + \frac{\tau}{N} (\sum_{k \neq i} \EX\left[x_i x_k \right] + \sum_{k \neq i} \EX\left[x_i x_k\right]) \text{if $i = j$ ,} \\ 
2 ( \mu - \frac{N-1}{N} \tau) \EX\left[x_i x_j\right] + \frac{\tau}{N} (\sum_{k \neq i} \EX\left[x_k x_j \right] + \sum_{k \neq j} \EX\left[x_k x_i\right] ),  \text{otherwise.} 
\end{cases}
\label{eq:rgbm-2nd-moment-dynamics}
\end{aligned}
\end{equation}
In the case of initial conditions that are identically distributed, it follows that for all $i$ $\EX\left[x_i^2(t)\right] = v(t)$ and for all pairs $i$ and $j$, with $i \neq j$, $\EX\left[x_i(t) x_j(t) \right] = r(t)$. Thus, the system of equations~(\ref{eq:rgbm-2nd-moment-dynamics}) reduces to 
\begin{align}
\frac{\mathrm{d}v}{\mathrm{d}t} &= 2\bigg(\mu - \frac{N-1}{N}\tau + \frac{\sigma^2}{2}\bigg) v  + 2 \frac{N-1}{N} \tau r, \label{eq:rgbm-2nd-moment-dynamics-reduced1}\\
\frac{\mathrm{d}r}{\mathrm{d}t} &= 2\bigg(\mu - \frac{1}{N}\tau\bigg) r  + \frac{2}{N} \tau v.
\label{eq:rgbm-2nd-moment-dynamics-reduced2}
\end{align}
This is a linear system whose solution can be explicitly found. Using the solution to the above system, and knowing that $\EX\left[x_i(t)\right] = x_0 \exp \left[ \mu t \right]$ , the relative variance can be rewritten as
\begin{align}
 \mathrm{R}_N(t) &= \frac{v(t) + (N-1) r(t)}{N x_0 \exp\left[2 \mu t\right]}-1.
    \label{eq:relative-variance-final}
\end{align}%
 
Hence, the critical self-averaging time $t_c$ can be phrased as the solution to
\begin{align}
 \frac{v(t_c) + (N-1) r(t_c)}{N x_0 \exp\left[2 \mu t_c\right]} = 2.   
\end{align}
 
\begin{figure*}[t!]
\includegraphics[width=12cm]{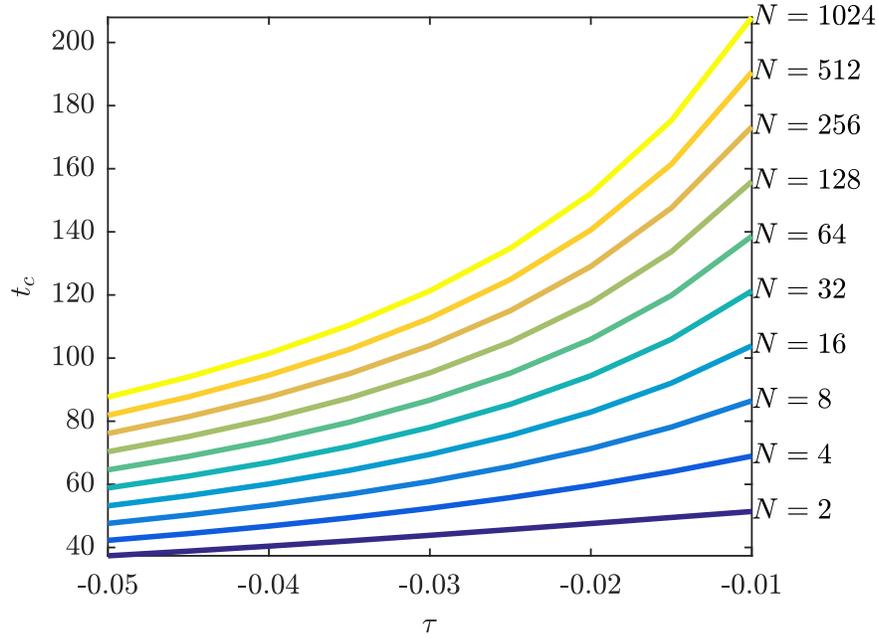}
\caption{Critical self-averaging time $t_c$. Critical self-average time $t_c$ as a function of $\tau$ for $\mu = 0.021$, $\sigma^2 = 0.02$  and varying $N$. \label{fig:critical-time}}
\end{figure*}

In the special case when $\tau = 0$, Eq.~(\ref{eq:relative-variance-final}), it follows that $v(t) = \exp\left[(2\mu +\sigma^2)t \right]$ and $r(t) = 0$. Thus, the relative variance reduces to
\begin{align*}
        R_N(t) &\propto \frac{\exp{\left[\sigma^2 t\right]} -1}{N},
\end{align*}
and the critical self-averaging time is 
\begin{align}
t_c = \frac{\log(N)}{\sigma^2}.
\label{eq:gbm-self-averaging}
\end{align}
This is the standard result for GBM and is known in the literature (See for example Ref.~\cite{peters2018sum}). Finding a general solution of Eq.~(\ref{eq:relative-variance-final}) in terms of $t_c$ for an arbitrary $\tau < 0$ is impossible. Only an implicit solution can be derived. This solution, for various population sizes $N$, is displayed in Fig.~\ref{fig:critical-time}. Obviously, the magnitude of $\tau$ affects non-linearly the critical self-averaging time. More importantly, more negative $\tau$ values also lead to lower $t_c$. %\textcolor{red}{ The inset in the figure gives the same plot, but now we fix $\tau = -0.025$ and vary $\sigma$. We observe that in RGBM, changes in the noise amplitude also exhibit a non-linear effect on the critical self-averaging time, which is significantly different from the GBM case (Eq.~(\ref{eq:gbm-self-averaging}). This is because in the relative variance equation there is a coupling term between $\tau$ and $\sigma$.\textbf{REMOVE THIS}}

\subsection{Magnitude of fluctuations}
We showed that after the self-averaging time, the population average wealth $\langle x(t) \rangle_N$ will randomly oscillate between positive and negative values. The fluctuations of the oscillations have upper and lower bounds, for respectively, their maximum and minimum values. The bounds are determined by the values of the RGBM parameters. To quantify the magnitude of these fluctuations we utilize Markov's inequality. The inequality states that
\begin{align}
    \mathrm{Pr}\bigg(|\langle x(t) \rangle_N| \geq h(t)\bigg) \leq \frac{\EX\left[\phi(|\langle x(t) \rangle_N|)\right]}{\phi(h(t))},
\end{align}
% where \textcolor{red}{$\phi: \mathbb{R}_{>0} \to \mathbb{R}_{>0}$ 
where $\phi(\cdot)$
is a monotonically increasing function for the non-negative reals. We set $\phi(x) = x^2$. Notice that in our setting $\phi(x)$ is even, i.e., $\phi(x) = \phi(-x) = \phi(|x|)$, which, as will be seen, greatly eases our analysis. For further simplicity, we are going to consider the situation where the initial condition is $x_0 = 1$. The results can be easily generalized to the case of arbitrary initial conditions.
 
From Eq.~(\ref{eq:relative-variance-final}) we know that
 \begin{equation}
 \begin{aligned}
  \EX\left[\phi(\langle x(t) \rangle_N \right)] &=  N v(t) + N (N-1) r(t) \\ 
  &= \frac{\exp\left[(2\mu-\tau+\frac{\sigma^2}{2})t\right]}{\mathrm{C}} \times \bigg(\mathrm{C} \cosh{\left[\frac{\mathrm{C} t}{2\sqrt{N}}\right]} - \sqrt{N} \big( \sigma^2 - 2\tau\big) \sinh{\left[\frac{\mathrm{C} t}{2\sqrt{N}}\right]}\bigg), \label{eq:second-moment}
 \end{aligned}
\end{equation}
where  $\mathrm{C} = \sqrt{N(\sigma^2 - 2 \tau)^2 + 8 \sigma^2 \tau}$. In the above equation, we substituted the exact values for $v(t)$ and $r(t)$, which can be found by solving the system given with Eqs~\eqref{eq:rgbm-2nd-moment-dynamics-reduced1} and~\eqref{eq:rgbm-2nd-moment-dynamics-reduced2}. 

It follows that,
\begin{equation}
\begin{aligned}
      \EX\left[\phi(\langle x(t) \rangle_N \right] &\leq \exp\left[(2\mu-\tau+\frac{\sigma^2}{2})t\right] \cosh\left[\frac{\mathrm{C}t}{2\sqrt{N}}\right] \\
      &\approx \exp\left[(2\mu-\tau+\frac{\sigma^2}{2} + \frac{\mathrm{C}}{2\sqrt{N}})t\right] \\
      &\leq \exp\left[2(\mu-\tau+\frac{\sigma^2}{2})t\right].
\end{aligned}
\end{equation}
Setting $h_\delta(t) = \exp\left[\delta(\mu-\tau+\frac{\sigma^2}{2})t\right]$, where $\delta > 0$ is an arbitrary constant, we get that
\begin{align} 
        \mathrm{Pr}\bigg(|\langle x(t) \rangle_N |\geq h_{\delta}(t)\bigg) &\leq \exp\left[(1-\delta)2(\mu-\tau+\frac{\sigma^2}{2})t\right]
\end{align}
which implies convergence in probability whenever $\delta >1$, i.e. for any $\eta >0 $, $\delta >1$, as $t \to \infty$

\begin{align}
\mathrm{Pr}\bigg(
 \frac{ |\langle x(t) \rangle_N|}{h_\delta(t)} \ge \eta  \bigg) \longrightarrow  0. 
\end{align}
This implies that $ \exp\left[(\mu-\tau+\frac{\sigma^2}{2})t\right]$ is an upper bound for the maximum average wealth and $ -\exp\left[(\mu-\tau+\frac{\sigma^2}{2})t\right]$ is a lower bound for the minimum average wealth in the population. 

Interestingly, the bounds indicate that there might be circumstances in which the observed time-averaged growth in a negative reallocation economy is larger than the one predicted by the ensemble average. However, we again restate the finding that there might also be time periods when the wealth is negative, and the time-average growth rate is undefined.

%As an application of the Borel-Cantelli lemma we can also conclude that we have an ever stronger result. Namely, for any $\delta > 1$ this leads to
%\begin{align}
%\mathrm{Pr}\bigg( \limsup_{n \to \infty}
% \frac{ |\langle x(t_n) \rangle_N|}{h_\delta(t_n)} = 0  \bigg) &= 1
%\end{align}
%for any sequence of positive reals $(t_n)_{n \in \mathbb{N}}$ that grows sufficiently rapidly. For example, it is enough that we have $t_n \sim_\infty  \sqrt[c]{n}$ for any fixed constant $c>0$.

\section{Discussion}
\label{sec:discussion}
We studied the non-ergodicity of the population average wealth in RGBM, a baseline model for a growing and reallocating economy. We found out that non-ergodicity is manifested in the difference between the time-average growth rate and the ensemble growth rate of the observable. Identically to the standard case of GBM, in srGBM the ensemble growth rate has exponential growth with a rate equal to the drift. The time-average growth rate, however, is non-existent as long as the reallocation rate is negative.

The existence of negative average population wealth in RGBM is rather counter-intuitive. We believe that this observation can be explained through real world factors that are not captured by RGBM. For instance, one such factor is debts to other generations~\cite{houle2014generation}. Obviously, this can be observed only in models that account for intergenerational interactions. Another factor is debts towards other economies~\cite{poirson2002external}, which are also excluded from the model. Our analysis reveals that negative average population wealth can arise in a closed one-generation economy simply as a consequence of non-ergodicity. To understand this result, let us look at the deterministic version of RGBM, i.e. the situation when $\sigma = 0$ in equation~(\ref{eq:rgbm}). In this case, we end up with an $N$-dimensional system of linear differential equations. Its $i$-th solution reads
\begin{equation}
\begin{aligned}
    x_i(t) = &\langle x(0) \rangle_N \exp{\left[ \mu t\right]} + \big(x_i(0) - \langle x(0) \rangle_N \big) \exp{\left[ (\mu-\tau) t\right]}.
    \label{eq:rgbm-deterministic-individual-solution}
\end{aligned}
\end{equation}
Obviously, the deterministic model is ergodic because it displays no randomness. The population average is always  $\langle x(0) \rangle_N \exp{\left[ \mu t\right]}$.
We showed that in the stochastic model, after self-averaging the randomness makes the population average to be dominated by extreme values. Due to the negativity of $\tau$, the extremes can be both positive and negative. This drives the system out of equilibrium even after rescaling.

RGBM is a continuous stochastic process and the observed oscillations within the range of the magnitude of fluctuations imply that at some point the average is zero. At this point, the rescaled wealth is undefined. We argue that this questions the validity of rescaling wealth by the population average in a non-ergodic economy, at least after the point of self-averaging. More importantly, in this case, the majority of the standard measures for economic well-being (such as inequality) are also undefined, since they are based on the assumption of positive population average wealth. Thus it can be argued that policies based on their dynamics cannot be conducted in a non-ergodic economy . In fact, the exact dynamical behavior of economic well-being measures in a non-ergodic economy is an ongoing debate~\cite{stojkoski2021income}. We believe that the results presented here can represent a starting point for resolving this issue.

\section*{Acknowledgements}
VS acknowledges financial support by the German Science Foundation (DFG, Grant number ME 1535/12-1).

\bibliographystyle{unsrt}
%\bibliography{mix-gbm}

\end{document}